\documentclass[fleqn,usenatbib]{mnras}

\usepackage[T1]{fontenc}
\usepackage{ae,aecompl}

\usepackage{graphicx}	
\usepackage{amsmath}	
\usepackage{amssymb}	

\title[WD - MS collisions from triple systems] {White dwarf - main sequence star collisions \\ from wide triples in the field }

\author[E. Michaely and M. M. Shara]{
Erez Michaely$^{1}$\thanks{E-mail: erezmichaely@gmail.com},
Michael M. Shara$^{2}$
\\
$^{1}$Astronomy Department, University of Maryland, College Park, MD 20742\\
$^{2}$Department of Astrophysics, American Museum of Natural History, Central Park West at 79th Street, New York, NY 10024, USA}

\date{Accepted XXX. Received YYY; in original form ZZZ}

\pubyear{2020}
\begin{document}
\label{firstpage}
\pagerange{\pageref{firstpage}--\pageref{lastpage}}
\maketitle

\begin{abstract}
Multiple star systems interact strongly with Galactic field stars when the outer semi-major axis of a triple or multiple star is $>10^3$ AU.
Stable triples composed of two white-dwarfs (WD) and a low mass main sequence (MS) star in a wide outer orbit can thus be
destabilized by gravitational interactions with random field stars. Such interactions excite the eccentricity of the distant third star sufficiently so that it begins to interact significantly with the inner binary. When this occurs the triple undergoes multiple binary-single resonant encounters. These encounters may result either in a collision between the nondegenerate component and a WD,
or the breakup of the triple into a compact binary and a third object which is ejected. The compact binary can be either a MS-WD pair which survives, or collides, or a double WD which may inspiral through gravitational wave emission. We calculate the collision rate between a MS and WD star, and the merger rate of double WDs. Additionally, we describe the prospects of detectability of such a collision, which may resemble a sub-luminous SN event.

\end{abstract}

\section{Introduction}
\label{sec:Introduction}
In recent years it had been shown that for ultra-wide stellar systems
the host galactic field is collisional \citep{Kaib2014,Antognini2016,Michaely2016,Michaely2019,Michaely2020,Michaely2020a}, 
that is one cannot regard the system as isolated from its environment. 
Therefore, when considering the evolution of wide stellar systems one needs
to account for random gravitational interactions with passing stars,
including close flybys, even in low density environments like the field of the host
galaxy. In turn, the interactions between flybys and wide systems
change the wide system's orbital characteristics, both the semi-major
axis (SMA) and its orbital eccentricity. The change in eccentricity
is usually the more pronounced effect. \citep{Lightman1977,Merritt2013a}.

Here we consider ultra wide systems to be either wide binaries or
wide hierarchical triples. \citet{Kaib2014} calculated the rate of
stellar collision in the Milky-Way (MW) Galaxy, caused by wide
(SMA > $1000 {\rm  AU}$) binaries interacting with random flyby stars in
the field. They found that a stellar collision between two main sequence
(MS) stars happens every $1000-7500 {\rm yrs}$ in the MW. Expanding
on that work \citet{Michaely2016} described the formation of low-mass
X-ray binaries (LMXBs) from wide binary systems composed of a stellar
compact object, i.e. a black-hole (BH) or a neutron star (NS) and a
low mass companion. They found that if BHs are born with little
to no natal kicks then the formation rate of LMXBs is consistent with the inferred
rate from observations. Additionally, \citet{Michaely2019} found
that gravitational waves (GW) sources are also formed from wide binary
BHs interacting with flyby stars in their host galaxies and reported a 
rate of $1-10 {\rm Gpc^{-3}yr^{-1}}$.

Hierarchical triples are composed of an inner binary and an outer binary,
where the inner binary center of mass (COM) is consider one component
of the outer binary and the distant, third object is the second component.
\citet{Michaely2020} calculated the GW merger rate for wide BH triples
that become unstable due to interactions with random field stars that
excite their outer eccentricity. As a result the triple becomes unstable
and a GW merger can happen either in the instability phase or at the
endstate phase. Under their assumptions they found an extremely high
merger rate of $100-250 {\rm Gpc^{-3}yr^{-1}}$, which indicates that
while the formation channel is robust the underlying assumptions
are very generous. Recently \citep[hereafter: Paper I]{Michaely2020a}
calculated via similar dynamical processes the Type Ia supernova (SN)
rate of triple white-dwarf (WD) systems. The SN originate either by
a direct collision from double WDs \citep{Raskin2009,Rosswog2009,Thompson2011,Kushnir2013} or by a GW inspiral that leads
to double WDs coalescing in a Hubble time, a process called ``the double-degenerate
(DD) channel" \citep{Iben1984,Webbink1984}.

Recently several studies were done on the dynamics of scattering events trying to calculate, among others, 
the rates of DWD mergers or collisions as a source for Type Ia SNe. \citet{Antognini2016} 
performed an extensive numerical study on the nature of scattering events: binary-binary, triple-binary, and triple-single (the focus of this manuscript).
In \citet{Antognini2016} only triples with outer SMA $<1000 \rm {AU}$ were considered, unlike paper I and this manuscript.
They reported that the collision rates of DWD in the field is  $\sim2\times10^{-6}\rm yr^{-1}$.
Population synthesis studies were carried out in order to quantify the Type Ia SNe originating from triples, due to secular evolution \citet{Hamers2013,Toonen2018}.
They showed that the fine tuning of the mutual orbital inclination combined with stellar evolution lower SN rates down to $0.1 - 1\%$ of the total rate.
Additionally, \citet{Hamers2018a} calculated the Type Ia SN originating from quadruple star systems, due to mergers and collisions. It was found the combined rate 
for all the possible channels that lead to a SN is 3 order of magnitude lower than the observed rate.
In \citet{Leigh2018a} it was shown that gravitational scattering is dominated by triple-single scattering events, 
similar to paper I and this manuscript, rather than binary-binary or triple-binary scattering.

In this study we follow-up the work of paper I and relax the assumption
that all three components of the triples are WDs. Instead we focus
on triples wherein two of the components are WDs and the third is a low
mass MS star. Following a similar dynamical scenario as explored in paper I, we
focus both on the intermediate phase, with
direct collisions between the components, and the endstate where we
calculate the rates where collisions are possible and the inspiral
of double WD (DWD) binaries due to GW emission.

The outcomes of a collision between a WD and a low mass MS star have been explored
in the literature \citep{Shara1977,Shara1978,Shara1986,Regev1987,Ruffert1992,Shara1999}.
Other stellar collisions have been studied as well \citep{Soker2006,Katz2012,Aznar-Siguan2013}.
Here we consider the energetics, timescales and the potential detectability of the expected
transient generated by a WD-MS collision and reserve a detailed study
for future research.

In section \ref{sec:Wide-triples-in}
we describe the interaction of ultra-wide triples in the field
of a host galaxy. In section \ref{sec:Binary-single} we discuss
the dynamics in the unstable phase of the triple as multiple binary-single
encounters. In section \ref{sec:Galactic-rates} we calculate
the expected rates of collisions and DD mergers in large spiral and elliptical galaxies. We discuss
the energetics and detectability of WD-MS collisions in section \ref{sec:WD-MS-collision}.
We estimate the delay-time distribution of WD-MS collisions in section \ref{subsec:Delay-time-distribution}. 
We summarize our results in section \ref{sec:Discussion-and-summary}.

\section{Wide triples in the field}
\label{sec:Wide-triples-in}
\begin{figure}

\includegraphics[width=1\columnwidth]{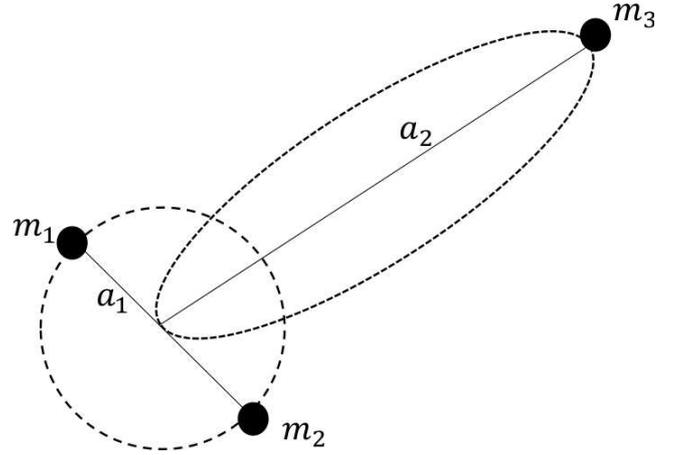}\caption{\label{fig:Illustration}Illustration of a triple system in hierarchical
configuration, $a_{1}\ll a_{2}$. The inner binary is circular with SMA, $a_1$. The outer binary is defined by the outer SMA, $a_2$ and outer eccentricity, $e_2$.}

\end{figure}

In what follows we briefly describe the gravitational interaction
between a wide hierarchical system, illustrated in Fig. \ref{fig:Illustration},
and a flyby star in the field of the host galaxy. A complete mathematical
description can be found in paper I and references therein.

For simplicity we assume throughout this paper that all three components
in the triple have the same mass $m_{1}=m_{2}=m_{3}=0.6M_{\odot}.$
The inner binary (masses $m_{1}$ and $m_{2}$) of the triple is characterized
by a circular orbit with a SMA $a_{1}$. The outer binary, composed
of the third object $m_{3}$ and the center of mass (COM) of the inner
binary is characterized by a SMA $a_{2}>1000 {\rm AU}$ and an eccentricity
$e_{2}$. The nature of the interaction between the triple and a random
flyby star is determined by the local environment of the host galaxy. In particular,
the encounter is determined by the relative speed between the flyby star and the triple, $v_{{\rm enc}}$
and the interaction rate, $f$. We identify the relative speed to
be the local velocity dispersion, $v_{{\rm enc}}=\sigma$. The interaction
rate is related to the local number stellar density $n_{*}$ by $f=n_{*}\sigma_{{\rm CS}}v_{{\rm enc}}$
where $\sigma_{{\rm CS}}$ is the geometrical cross-section of the flyby interaction.

In paper I it was shown that for an ensemble of wide triple systems
with a thermal distribution, $f\left(e_{2}\right)=2e_{2}$ of outer
eccentricities, there is a non-negligible probability that the third
component's pericenter passage, $q=a_{2}\left(1-e_{2}\right)$, is
sufficiently close to the inner binary COM that the systems becomes
unstable. Specifically, the instability condition is $q\le a_{1}$.

The instability of the triple can be described as follows.
The triple undergoes multiple binary-single encounters, during which
a temporary binary is formed from a random pair of the three system
components, with a SMA $a_{{\rm IMS}}$ and eccentricity $e_{{\rm IMS}}$,
where the subindex IMS stands for ``intermediate state". The values of $a_{{\rm IMS}}$
are drawn uniformly between $\left(a_{1},2a_{1}\right)$ and the $e_{{\rm IMS}}$
are drawn from a thermal distribution (for a complete derivation see
eq. 20-24 in paper I). The third object is bound to the temporary binary
on a wide Keplerian orbit with $a_{2{\rm IMS}}$ determined by conservation
of orbital energy. The relatively wide outer orbit allows the inner
binary to revolve around its COM multiple times before the next pericenter
passage, namely before the next binary-single interaction where this
process repeats.

For point-like particles the average number of binary-single encounters
is $\left\langle N\right\rangle =20$ \citep{Michaely2020}. The end
state is the formation of a compact binary with SMA $a_{{\rm ES}}<a_{1}$,
and eccentricity $e_{{\rm ES}}$ which follows a thermal distribution.
The end state binary is composed of two objects randomly chosen out of the
triple. The third object is ejected to infinity.

In order to determine the rate at which wide triple systems become unstable
we focus on the outer binary of the wide triple, with $a_{2}$ and
$e_{2}$. This wide binary interacts with flyby stars, of mass $m_{{\rm p}}$,
which perturb the outer binary. \citet{Lightman1977} and \citet{Merritt2004}
showed that the perturbation primarily torques the outer binary,
changing its eccentricity. We use the ``loss-cone" formalism to
determine the loss rate of wide stable triple star systems. 

We consider an ensemble of wide systems with a thermal distribution of the outer eccentricity. The loss cone is the fraction of systems from the ensemble
that becomes unstable, namely the outer pericenter distance of the outer binary is within the inner binary SMA, $q\le a_{1}$. From paper I we get:
\begin{equation}
F_{q}=\frac{2a_{1}}{a_{2}}
\end{equation}
which for hierarchical triples $F_{q}\ll1$. We compare this value
to the smear-cone, $F_{s}$ (paper I) which essentially represents the change
in eccentricity due to the impulse interaction with the perturber.
\begin{equation}
F_{s}=\frac{27}{4}\left(\frac{m_{p}}{M}\right)^{2}\left(\frac{GM}{a_{2}v_{{\rm enc}}^{2}}\right)\left(\frac{a_{2}}{b}\right)^{4}
\end{equation}
where $M=m_{1}+m_{2}+m_{3}$ is the total mass of the triple and $b$
is the closest approach of the perturber to the triple COM. These
quantities naturally create two regimes, the full loss cone and the
empty loss cone regimes. In the full loss cone regime, where $F_{s}\ge F_{q}$
the loss cone is continuously full, namely the interactions are frequent
and strong enough to keep the loss cone full. In the empty loss
cone regime, $F_{s}<F_{q}$ the loss cone is primarily empty until
a weak interaction with a flyby star occurs which kicks a system into the loss
cone.

Additionally, we account for binary ionization processes in the field.
Because the field is collisional for wide binaries we need
to account for the loss of wide binaries due to the disruption of
wide binaries from random gravitational processes. We use the standard
half-life calculation from \citep{Bahcall1985}
\begin{equation}
t_{1/2}=0.00233\frac{v_{{\rm enc}}}{Gm_{{\rm p}}n_{*}a_{2}}.
\end{equation}
The loss probabilities, the probability of a system with inner SMA $a_1$, outer SMA $a_s$, located in a stellar environment with number density $n_*$
at time $t$ to become unstable, are given by the following equations: for the
empty loss cone regime 
\begin{equation}
L\left(a_{1},a_{2},n_{*}\right)_{{\rm empty}}=\tau\frac{2a_{1}}{a_{2}}n_{*}\pi\sqrt{\frac{27}{8}\left(\frac{m_{{\rm p}}}{M}\right)^{2}\frac{GMa_{2}^{4}}{a_{1}}}\left(1-e^{-t/\tau}\right)\label{eq:L_empty}
\end{equation}
where $\tau=t_{1/2}/\ln2$. For the full loss cone regime we get 
\begin{equation}
L\left(a_{1},a_{2},n_{*}\right)_{{\rm full}}=\tau\frac{2a_{1}}{a_{2}}\left(\frac{GM}{4\pi^{2}a_{2}^{3}}\right)^{1/2}\left(1-e^{-t/\tau}\right).\label{eq:L_full}
\end{equation}

\section{Binary-single encounters}
\label{sec:Binary-single}
In this section we focus on the dynamics of the triple system when
it becomes unstable and enters the multiple binary-single interaction
phase. In subsection \ref{subsec:The-intermediate-phase} we focus
on the intermediate phase while in subsection \ref{subsec:End-state}
we explore the end state of the intermediate phase, specifically the
compact binary that forms as a result of the intermediate phase.

The case where one of the components is a low mass star, in its MS
phase, is fundamentally different from the case presented in paper
I and in \citet{Michaely2020}. The reason is that the radius of a low
mass MS star is roughly $50$ times larger than the radius of a WD. 
Hence a direct collision between a WD and the MS star becomes probable.

\subsection{The intermediate phase}
\label{subsec:The-intermediate-phase}
In order to calculate the probability of this scenario, a collision
between a MS star and a WD from a triple with two WDs and one MS, star we
perform a numerical calculation. We sample 20 values of the inner
SMA $a_{1}$ from $\left(10^{-2}{\rm AU},10^{2}{\rm AU}\right)$ equally
distributed in log space. For each value of $a_{1}$ we perform $N_{{\rm tot}}=10^{5}$
``scattering experiments".
In this context a single ``scattering experiment" is $N_{{\rm IMS}}=20$
binary-single interactions, where for each one a temporary binary is formed
with $a_{{\rm IMS}}$ and $e_{{\rm IMS}}$. $a_{{\rm IMS}}$ is drawn
uniformly from $\left(a'_{{\rm L}},a'_{{\rm U}}\right)$ and $e_{{\rm IMS}}$
is drawn from a thermal distribution. The boundary values $\left(a'_{{\rm L}},a'_{{\rm U}}\right)$ are determined by the individual masses of the triple (see eq. 18-24 in paper I).
Moreover, each temporary binary is composed of two random components, such that either a double
WD is formed or a WD-MS binary is formed. The temporary pericenter
is $q'=a_{{\rm IMS}}\left(1-e_{{\rm IMS}}\right)$ and we compare
it to the sum of the radii of the temporary binary, $R=r_{1}+r_{2}$.
We calculate the radius of the WD using \citet{Hamada1961}:
\begin{equation}
R_{{\rm WD}}=1.3\times10^{-2}R_{\odot}\text{\ensuremath{\left(\frac{M_{{\rm WD}}}{0.6M_{\odot}}\right)}}^{-1/3},
\end{equation}
and for the low mass convective MS star we use 
\begin{equation}
R_{*}=0.6R_{\odot}\left(\frac{m_{*}}{0.6M_{\odot}}\right)
\end{equation}
taken from \citet{Mann2015}.

where $m_{*}$ is the mass of the MS star. In the case where $q'\le R$
we flag the scattering experiment as a collision, check the nature
of the components that collide and terminate this experiment. In the
case where $q'>R$ we calculate the outer binary orbital period, which
is the time of the next interaction and randomize the binary-single
interaction again until we reach $N_{{\rm IMS}}$ times.

\begin{figure}
\includegraphics[width=1.05\columnwidth]{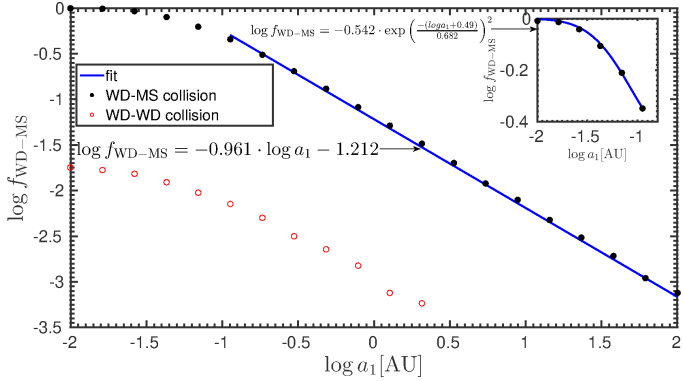}\caption{\label{fig:f_coll_IMS}The fraction of systems that undergo a WD-MS
collision in the intermediate state. $f_{{\rm WD-MS}}$ as a function
of the initial SMA, $a_{1}$. For every initial value of $a_{1}$
we simulated $10^{5}$ scattering experiments. In each experiment we
run $N_{{\rm IMS}}=20$ binary-single scattering events. For every
event we randomize the temporary binary orbital, SMA and eccentricity.
We check if the temporary IMS leads to a WD-MS collision or WD-WD
collision. Black dots are the calculated fraction of WD-MS collisions
from our numerical experiment. Red circles are the fraction of WD-WD
collisions. The blue solid line is a broken fit to the numerical results.}

\end{figure}

In order to calculate the fraction of systems that undergo a collision
between a WD and a MS star in the intermediate phase, $f_{{\rm WD-MS}}\left(a_{1}\right)$
we divided the number of mergers by $N_{{\rm tot}}$ for each $a_{1}$.
The numerical results are presented in Fig. \ref{fig:f_coll_IMS}. We
fit the numerical results with two functions as a function of the
inner SMA $a_{1}$.
\begin{equation}
\log f_{{\rm WD-MS}}\left(a_{1}\right)=\begin{cases}
-0.961\cdot\log a_{1}-1.212 & a_{1}>0.1{\rm AU}\\
-0.542\cdot\exp\left(-\frac{\left(\log a_{1}+0.49\right)}{0.682}\right)^{2} & a_{1}<0.1{\rm AU}
\end{cases}.\label{eq:IMS_Collision}
\end{equation}

We note here that we ignore the rate of double WD collisions
in the intermediate stage, represented in Fig. \ref{fig:f_coll_IMS}
with red circles. One can expect the rate of double WD collision scales
as the ratio of the radii of WD and the MS star, for $M_{{\rm WD}}=m_{*}=0.6M_{\odot}$
\begin{equation}
\frac{R_{{\rm WD}}}{R_{*}}\approx0.02.
\end{equation}

\subsection{End state}
\label{subsec:End-state}
\begin{figure*}
\includegraphics[width=1.0\columnwidth]{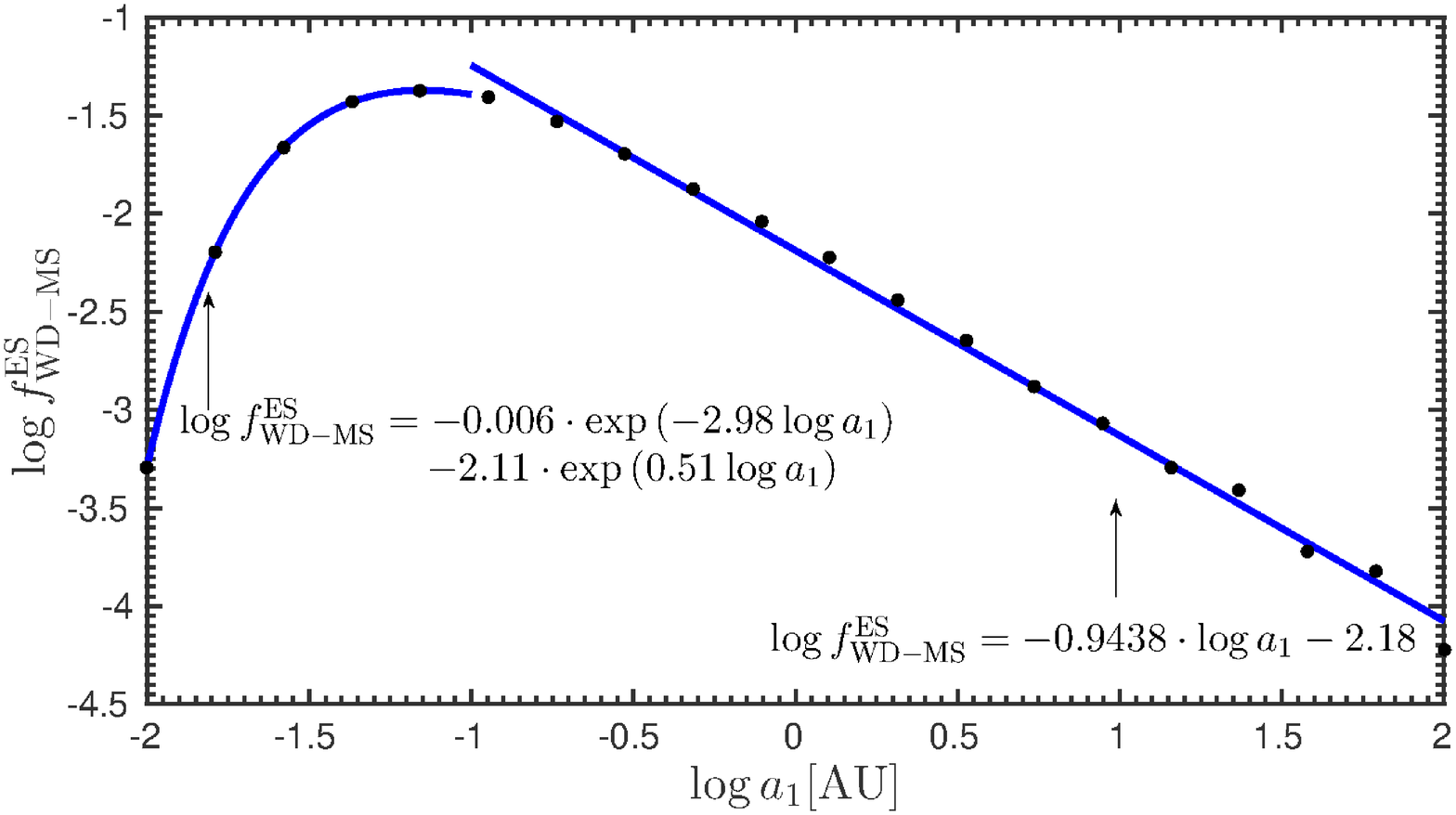} $\quad$ \includegraphics[width=1.0\columnwidth]{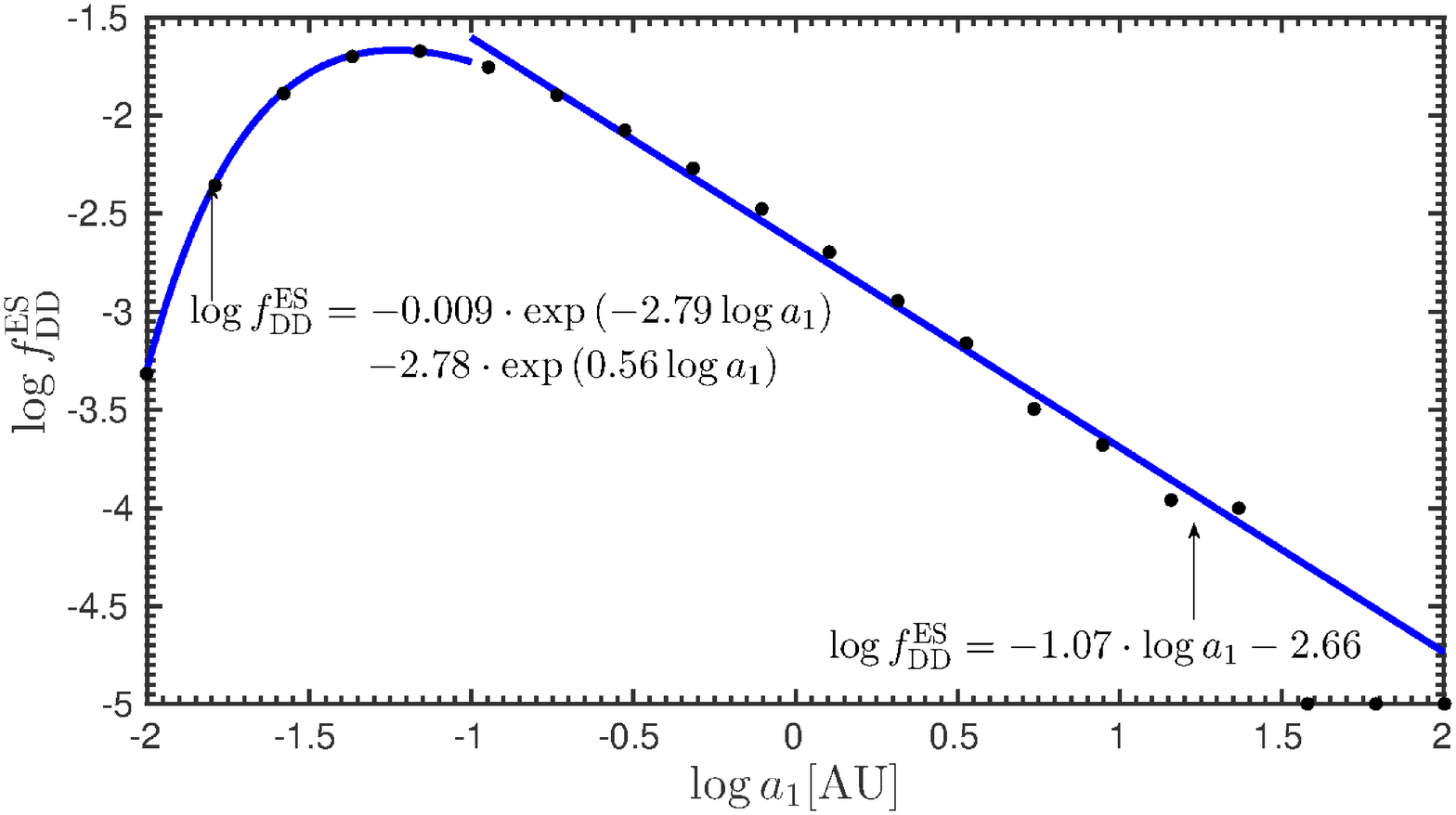}\caption{\label{fig:ES}Left plot: Similar to Fig. \ref{fig:f_coll_IMS}, but
the fraction of systems that undergo a WD-MS collision in the endstate.
$f_{{\rm WD-MS}}^{{\rm ES}}$ as a function of the initial SMA, $a_{1}$.
If the triple survived the intermediate state, the final outcome is
a compact binary with SMA drawn from eq. (\ref{eq:E_final}) and the
eccentricity is drawn from a thermal distribution. We check if a compact
binary leads to a WD-MS collision. Black dots are the calculated fraction
of WD-MS collisions from our numerical experiments. Blue solid line is
a broken fit to the numerical results. Right plot: Same as left
plot but we calculate the fraction of DWD systems that merge via GW
emission within a Hubble time.}

\end{figure*}

In the case where no collision occurs during the $N_{{\rm IMS}}$
times we simulate the endstate configuration. The emerging compact
binary has a different SMA distribution than the distribution in the
intermediate phase. Specifically, the energy of the final binary is
distributed according to the following relation \citep{Stone2019b}
\begin{equation}
E_{{\rm ES}}\propto\left|E_{1}\right|^{-4}.\label{eq:E_final}
\end{equation}
where $E_{{\rm ES}}$ is the energy of the endstate binary and $E_{1}=-Gm_{1}m_{2}/(2a_{1})$
is the initial binary energy. Similar to the intermediate phase case
the eccentricity of the compact binary $e_{{\rm ES}}$ is drawn from a
thermal distribution \citep{Stone2019b}. Hence, a compact binary
is formed for every system that did not collide during the IMS with
a SMA $a_{{\rm ES}}\left(E_{{\rm ES}}\right)$ and eccentricity $e_{{\rm ES}}$.
This endstate binary is composed either of a double WD (DWD) or a
MS star with a WD companion. As a result the possible outcomes
are either a direct collision between the components or the ensuing evolution of
a binary star. Depending on the binary components, the evolution may
lead to a merger through gravitational waves in the case of a DWD,
or any standard outcome allowed by binary evolution theory, including common
envelope evolution (CEE), a cataclysmic variable (CV) etc. In the left plot of
Fig. \ref{fig:ES} we present the fraction of systems that experienced
a WD-MS collision in the post-resonance phase as a function of the
initial SMA. The numerical fit presented in the figure is composed
of two functions:  

\begin{equation}
\log f_{{\rm WD-MS}}^{{\rm ES}}\left(a_{1}\right)=\begin{cases}
-0.94\cdot\log a_{1}-2.18 & a_{1}>0.1{\rm AU}
\\-0.006\cdot\exp\left(-2.98\log a_{1}\right) & a_{1}<0.1{\rm AU}\\
-2.113\cdot\exp\left(0.51\log a_{1}\right)

\end{cases}.\label{eq:ES_collsion}
\end{equation}

Next, we calculate the fraction of systems that undergo a double WD
merger through GW emission similar to the classic DD scenario. In
order to do so we calculate the GW merger timescale for eccentric
binaries which is \citep{Pet64} 
\begin{equation}
t_{{\rm merger}}\approx\frac{768}{425}T_{c}\text{\ensuremath{\left(a\right)}}\left(1-e^{2}\right)^{7/2}\label{eq:t_merger}
\end{equation}
where $T_{c}=a^{4}/\beta$ is the merger timescale for a circular
orbit and $\beta=64G^{3}m_{i}m_{j}\left(m_{i}+m_{j}\right)/\left(5c^{2}\right)$.
The indices $i,j$ are the indices of the two WDs that ended up as
the surviving compact binary and $c$ is the speed of light. If $t_{{\rm merger}}<10^{10}{\rm yr}$
we flag this systems as a DD inspiral, and a possible source for a
Type Ia SN. The numerical fit is
\begin{equation}
\log f_{{\rm DD}}\left(a_{1}\right)=\begin{cases}
-1.04\cdot\log a_{1}-2.647 & a_{1}>0.1{\rm AU}\\
-0.009\cdot\exp\left(-2.79\log a_{1}\right) & a_{1}<0.1{\rm AU}\\-2.77\cdot\exp\left(0.56\log a_{1}\right). 
\end{cases}\label{eq:ES_DD}
\end{equation}

We emphasize here that the outcome of a WD-MS binary that does not collide
is still of interest. These binaries are the progenitors of CEE 
or CV. Once the MS star companion in the binary evolves to the giant phase
it may fill its Roche-lobe and begin mass transfer. If the mass
transfer is stable an accretion disk forms around the WD and a CV
is formed. In the case where the mass transfer is unstable the system
may inspiral to a CEE. The study of the evolution of these
binaries has been explored elsewhere (for a recent overview see \citet{Beccari2019}).

\section{Galactic rates}
\label{sec:Galactic-rates}
In what follows we use the numerical fits found in the previous section
in order to compute rates in galaxies for WD-MS collisions in the
intermediate phase and in the endstate. Additionally, we compute the
galactic rates of DD inspiral.

\subsection{Galaxy models}
\label{subsec:Galaxy-models}
As described in section \ref{sec:Wide-triples-in}, the process that
destabilizes wide triples depends on the local stellar environment.
Specifically these are  the local stellar density, $n_{*}$ and the encounter
velocity, $v_{{\rm vec}}$ which is determined by the local velocity
dispersion. Hence, one needs to model these properties of the host
galaxy.

First, we model the MW Galaxy as a prototype of a large spiral galaxies
in the local universe, using the same model described in paper I
which follows \citep{Juric2008}. The number stellar density is given
by

\begin{equation}
n_{{\rm *s}}\left(r\right)=n_{0}e^{-\left(r-r_{\odot}\right)/R_{l}}\label{eq:MW_galaxy-1}
\end{equation}
where $n_{{\rm *s}}$ represents the number stellar density for and
$n_{0}=0.1 {\rm pc}^{-3}$ is the number stellar density in the solar
neighborhood, $R_{l}=2.6 {\rm kpc}$ is the Galactic length scale and
$r_{\odot}=8 {\rm kpc}$ is the distance of the Sun from the Galactic
center. The velocity dispersion we use is that 
of the flat rotation curve of the Galaxy, i.e. $\sigma=50 {\rm kms^{-1}}$
which in turn is set to be the encounter velocity.

We define ${\rm dN_{s}\left(r\right)}$, the number
of stars in a volume of the disk at a distant $r$ from the center
of the Galaxy, by 
\begin{equation}
{\rm dN_{s}(r)}={\rm n_{*s}\left(r\right)\cdot2\pi\cdot r\cdot h\cdot dr}
\end{equation}
where $h=1 {\rm kpc}$ is the scale height of the disk.

Second, for simplicity we model an elliptical galaxy the same way we did in paper I following
\citet{Hernquist1990}. The density profile is given by
\begin{equation}
n_{{\rm *e}}\left(r\right)=\frac{M_{{\rm galaxy}}}{2\pi r}\frac{r_{*}}{\left(r+r_{*}\right)^{3}}\label{eq:elliptical-1}
\end{equation}
where $n_{{\rm *e}}$ is the stellar density for \textit{elliptical}
galaxy and $r_{*}=1 {\rm kpc}$ is the scale length of the galaxy,
$M_{{\rm galaxy}}=10^{11} M_{\odot}$ is the total stellar mass of
the galaxy. Hence, 
\begin{equation}
{\rm dN_{{\rm e}}\left(r\right)}=\frac{n_{{\rm *e}}}{\left\langle m\right\rangle }{\rm dV}
\end{equation}
is the number of stars within some local volume ${\rm dV}$ at a distance
$r$ from the center. $\left\langle m\right\rangle =0.6M_{\odot}$
represents the average stellar mass of the galaxy. The velocity dispersion
for a typical elliptical galaxy is $\sigma=160{\rm kms^{-1}}$ \citep{Cappellari2013}.
In both types of galaxies we set the mass of the perturber to
be $0.6M_{\odot}$, the average mass of a star in the galaxy.

\subsection{WD-MS collision rate}
\label{subsec:WD-MS-collision-rate}
In this section we calculate the WD-MS collision rate in the intermediate
stage and endstate in both model spiral and elliptical galaxies. To do this
we must estimate the fraction of triple system which host two
WDs and a MS star. There are two possibilities for such systems to
exist. The first is a  DWD as the inner binary and the stellar companion
is the tertiary; we term these systems as WWM. The second, is a WD-MS
system as an inner binary and additional WD as the tertiary; we term
these systems as WMW.

We start by estimating the fraction of WWMs out of the stellar population,
$f_{{\rm WWM}}.$ We assume all stars with mass in the range of $1M_{{\rm \odot}}-8M_{\odot}$
become WDs and we count only MS stars in the mass rage of $0.4M_{\odot}-1M_{\odot}$.
We do so to be consistent with our assumption that the triple systems is composed
of stars of equal masses. We use the initial mass function (IMF) given
by \citep{Kroupa2001} and find that $f_{{\rm primary}}\approx0.1$
of all stars are in the range of $1M_{{\rm \odot}}-8M_{\odot}$
which turn into WDs in $10 {\rm Gyrs}$. This is an upper limit because
binary evolution can hasten WD faster via interactions between
the binary stars. The binary companion mass is calculated from a uniform
mass ratio distribution \citep{Moe2016}, $Q_{{\rm inner}}\in\left(0.1,1\right)$.
For the tertiary we calculate the outer mass ratio by the following
expression, $Q_{{\rm outer}}=m_{3}/\left(m_{1}+m_{2}\right)$ and
its value is drawn from a power law distribution \citep{Moe2016}
$f_{Q_{{\rm outer}}}\propto Q_{{\rm outer}}^{-2}$ where $Q_{{\rm outer}}\in\left(0.1,1\right)$.
For WWM systems we count only the systems in which the primary mass is within
the range $1M_{{\rm \odot}}-8M_{\odot}$, the secondary is in the
range $1M_{{\rm \odot}}-8M_{\odot}$ and the tertiary is in the
range $0.4M_{\odot}-1M_{\odot}$. Given these distributions we
find that the fraction of secondaries in the WD production range
is $f_{{\rm secondary}}\approx0.44$, while for the tertiaries it is $f_{{\rm tertiary}}\approx0.44$.
The triple fraction is set to be $f_{{\rm triple}}=0.2$ \citep{Duchene2013}
and the fraction of wide outer binaries greater than $1000 {\rm AU}$
from a log-uniform distribution, $f_{a_{2}}$, is $f_{{\rm wide}}=0.2.$
Combining these estimates we get 
\begin{equation}
f_{{\rm WWM}}=f_{{\rm primary}}\times f_{{\rm secondary}}\times f_{{\rm tertiary}}\times f_{{\rm triple}}\times f_{{\rm wide}}\approx7.6\times10^{-4}.\label{eq:f_WWM}
\end{equation}
A similar calculation can be made for the WMW case. Then the primary
is a WD, $f_{{\rm primary}}\approx0.1$ the secondary is a MS star,
$f_{{\rm secondary}}\approx0.4$ and the tertiary is a WD, $f_{{\rm tertiary}}\approx0.14$.

\begin{equation}
f_{{\rm WMW}}\approx2.2\times10^{-4}.\label{eq:f_WMW}
\end{equation}
We define 
\begin{equation}
f_{{\rm model}}=f_{{\rm WWM}}+f_{{\rm WMW}}.
\end{equation}

We note that this approach to estimate the fraction of triples is
a simplification of a very complex calculation. In order to calculate
the wide triple systems out of a certain stellar population more accurately
one needs to numerically evolve large numbers of systems and take into account both single
and binary stellar evolution. This, in turn, will change the
initial SMA and eccentricity distributions while also changing the masses.
In particular, common envelope evolution \citep{Ivanova2013} modifies
the inner SMA and even the outer SMA due to mass loss from the inner
binary \citep{Michaely2019a,Igoshev2020}.

Next, we compute the total galactic rate for WD-MS collisions for
both types of galaxies. The rate, $\Gamma$ is given by integrating
the loss cone (\ref{eq:L_empty}) and (\ref{eq:L_full}) for all outer
SMAs, $a_{2}$ between $10^{3}-10^{5}{\rm AU}$, the local stellar
density in the galaxy $n_{*}$ from equations (\ref{eq:MW_galaxy-1})
and (\ref{eq:elliptical-1}). In order to integrate the inner binary
SMA we use the following limits $10^{-1}\left(10^{-2}\right)-10^{2}{\rm AU}$.
We choose two minimal values of the inner binary in order to roughly estimate
the uncertainties caused by our lack of knowledge of the real distribution:
\begin{equation}
\Gamma=\int\int\int\frac{L_{{\rm collision}}\left(a_{1},a_{2},n_{*}\right)}{10{\rm Gyr}}da_{1}da_{2}{\rm dN\left(r\right)}\label{eq:the_integral-1}
\end{equation}
where $L_{{\rm collision}}\equiv L\left(a_{1},a_{2},n_{*}\right)f_{a_{1}}f_{a_{2}}f_{{\rm model}}f_{{\rm WD-MS}}$
and we define 
\begin{equation}
{\rm dL}\equiv\frac{L_{{\rm collision}}\left(a_{1},a_{2},n_{*}\right)}{10{\rm Gyr}}da_{1}da_{2}{\rm dN\left(r\right)}.\label{eq:dL}
\end{equation}
Now we plug in the function from eq. (\ref{eq:IMS_Collision})
\[
\Gamma_{{\rm MW}}=\int_{{\rm 0.5kpc}}^{{\rm 15kpc}}\int_{{\rm 10^{3}AU}}^{10^{5}{\rm AU}}\int_{{\rm 10^{-1}\left(10^{-2}\right)AU}}^{10^{2}{\rm AU}}{\rm dL} 
\]
\begin{equation}
\approx1.9\ \left(3.69\right)\times10^{-4} {\rm yr^{-1}}
\end{equation}
and for a typical elliptical 
\[
\Gamma_{{\rm elliptical}}=\int_{{\rm 0.1kpc}}^{{\rm 30kpc}}\int_{{\rm 10^{3}AU}}^{10^{5}{\rm AU}}\int_{{\rm 10^{-1}\left(10^{-2}\right)AU}}^{10^{2}{\rm AU}}{\rm dL} 
\]
\begin{equation}
\approx3.44\ \left(6.34\right)\times10^{-4} {\rm yr^{-1}.}
\end{equation}
These results are averaged over a $10 {\rm Gyr}$ lifetime of the galaxies,
In subsection \ref{subsec:Delay-time-distribution} we approximate
the delay-time distribution of these collisions.

In the case where a direct collision did not occur during the IMS
phase the triple is disrupted and a compact binary is formed. We calculate
the collision rate by using, $\log f_{{\rm WD-MS}}^{{\rm ES}}$ from
(\ref{eq:ES_collsion}) to get for the MW-like Galaxy 
\[
\Gamma_{{\rm MW}}^{{\rm ES}}=\int_{{\rm 0.5kpc}}^{{\rm 15kpc}}\int_{{\rm 10^{3}AU}}^{10^{5}{\rm AU}}\int_{{\rm 10^{-1}\left(10^{-2}\right)AU}}^{10^{2}{\rm AU}}{\rm dL}
\]
\begin{equation}
\approx1.56\ \left(1.14\right)\times10^{-5}{\rm yr^{-1}}
\end{equation}
 and for the elliptical galaxy a rate of 
 \[
 \Gamma_{{\rm elliptical}}^{{\rm ES}}=\int_{{\rm 0.1kpc}}^{{\rm 30kpc}}\int_{{\rm 10^{3}AU}}^{10^{5}{\rm AU}}\int_{{\rm 10^{-1}\left(10^{-2}\right)AU}}^{10^{2}{\rm AU}}{\rm dL}   
 \]
\begin{equation}
 \approx2.81\ \left(2.00\right)\times10^{-5}{\rm yr^{-1}.}
\end{equation}

\subsection{Double degenerate inspiral rate}
\label{subsec:Double-degenerate-inspiral}
Similar to the previous subsection here we calculate the galactic
rate for an endstate binary WD inspiral the may lead to Type Ia SN. In
this case we use eq. (\ref{eq:ES_DD}) and insert it in eq. (\ref{eq:dL})
to get 
\[
\Gamma_{{\rm MW,DD}}^{{\rm ES}}=\int_{{\rm 0.5kpc}}^{{\rm 15kpc}}\int_{{\rm 10^{3}AU}}^{10^{5}{\rm AU}}\int_{{\rm 10^{-1}\left(10^{-2}\right)AU}}^{10^{2}{\rm AU}}{\rm dL} 
\]
\begin{equation}
\approx7.56\ \left(5.38\right)\times10^{-6}{\rm yr^{-1}}
\end{equation}
and for a typical elliptical galaxy 
\[
\Gamma_{{\rm elliptical,DD}}^{{\rm ES}}=\int_{{\rm 0.5kpc}}^{{\rm 15kpc}}\int_{{\rm 10^{3}AU}}^{10^{5}{\rm AU}}\int_{{\rm 10^{-1}\left(10^{-2}\right)AU}}^{10^{2}{\rm AU}}{\rm dL} 
\]

\begin{equation}
\approx1.34\ \left(0.96\right)\times10^{-5}{\rm yr^{-1}}.
\end{equation}
These rates are of order one percent of the
total Type Ia rate observed in large galaxies:$\sim$ $10^{-3} {\rm yr^{-1}}$ \citep{Maoz2014}.

\section{WD-MS collision}
\label{sec:WD-MS-collision}
In what follows we discuss the physics and detectability of a WD-MS
collision event. \citet{Shara1977} studied the timescales and energetics of WD-MS collisions,
while \citet{Shara1978,Shara1986,Regev1987} carried out 2D hydrodynamics simulations of
head-on collisions, including a simple power law prescription to allow for nuclear energy release.
In their study a direct collision between a MS star with mass $M_{*}$
and radius $R_{*}$ and a WD with mass $M_{{\rm WD}}$ and radius
$R_{{\rm WD}}$ was considered. The relative velocity, at the collision,
was set $v_{{\rm coll}}\sim v_{*{\rm esc}}$ to the escape velocity of
the MS star in \citet{Shara1986}, and to 2000 or 6000 $kms^{-1}$ in \citet{Regev1987}.

As the WD approaches the MS star tidal forces act on the MS and stretch
its outer envelope. Energy conversion (from kinetic to thermal) and nuclear energy lgeneration in the collision event begins as the mass from
the outer edges of the MS star impacts the surface of the WD, decelerating and being deflected. 
The collision time is approximated by
\begin{equation}
\tau_{{\rm col}}\approx\frac{R_{*}}{v_{{\rm coll}}}\label{eq:t_coll}
\end{equation}
For $M_{*}=M_{{\rm WD}}=0.6M_{\odot}$ and $R_{*}=0.6R_{\odot}$ this
yields, $\tau_{{\rm col}}\approx900{\rm sec}.$ During this time, during
which the WD moves supersonically through the envelope of the MS
star, a roughly spherical shock is formed when the collision velocity
is lower than the WD's escape velocity, $v_{{\rm esc,WD}}\approx7\times10^{3}{\rm kms^{-1}}.$ The
shock compresses the envelope by a factor of $\sim4-10$ to average
values of $\rho\approx10^{3}{\rm gr\cdot cm^{-3}}$ and heats it to
$\sim3-5\times10^{8}K$. At this point radiation pressure becomes
dominant over gas pressure in the MS star's envelope. Under these conditions
proton capture and the hot CNO cycle are the dominant energy sources in the star, 
and He burning may also be initiated. The nuclear
burning consumes a few percent of the available hydrogen and releases
$E\sim2\times10^{48} {\rm ergs}$, during $\tau_{{\rm col}}$ and deposits
it in the escaping and optically thick stellar envelope. The binding
energy of a $0.6M_{\odot}$ star is $E_{G}\approx10^{48} {\rm erg}$;
this implies that the MS is totally disrupted by the collision. For a 10 $M_{\odot}$
MS star \citet{Regev1987} the energy release approaches s $\approx10^{50}{\rm erg}$.

The envelope escapes with a velocity of order a few x$10^{3} {\rm kms^{-1}}$,
assuming adiabatic expansion during the time it takes the optically thick
envelope to become optically thin, namely $\rho\approx10^{-11}{\rm gr\cdot cm^{-3}}$
is about $t\approx10^{6}-10^{8} {\rm sec}$. Combining the time to
achieve an optically thick expanding shell, $t$ and the total energy released
from the nuclear burning, $E$ one can estimate the bolometric luminosity
to be
\begin{equation}
L\approx\frac{E}{t}\approx\frac{10^{48}{\rm erg}}{10^{6}-10^{8}{\rm sec}}=10^{7}-10^{9} L_{\odot}.
\end{equation}
This luminosity is similar to that of a kilonova, but intermediate and heavy elements are unlikely to be produced during 
such events unless helium burning is achieved.

We note that all of the quoted 2D simulations were based on head-on collisions with simplistic nuclear energy release models.
3D hydrodynamics, including a realistic nuclear reactions network and off-center collisions are essential 
to better understand the outcomes of WD-MS star collisions. We reserve such studies to future efforts, noting that a
head-on collision is far from the most representative case.

\section{Delay time distribution (DTD)}
\label{subsec:Delay-time-distribution}
In this subsection we calculate the expected delay-time distribution
(DTD) of the transients described in section \ref{sec:WD-MS-collision}.
The DTD is the hypothetical rate of transients that follow a brief
star formation episode. The main channel of WD-MS collisions originates
during the multiple binary-single encounter when the triple becomes
unstable. The collision occurs on a dynamical timescale which is extremely
short compared to the stellar evolution time that was needed to produce
two WDs. Therefore, we only calculate the time since the star formation event which
produced the WD and RD stars.

The DTD is determined by the numbers of available triples that become
unstable as a function of time. This depends on the initial
mass function and the stellar evolution time for each mass. As already assumed, 
the rate of flyby interaction is constant in time if one disregards binary ionization via other processes (such as mass loss during
binary evolution). Therefore we can write the following dependency
\begin{equation}
\frac{d{\scriptscriptstyle N}}{dt}\propto\frac{d{\scriptscriptstyle N}}{dm}\frac{dm}{dt}.
\end{equation}
The first term is the initial mass function which is similar
to those of Kroupa and Salpeter \citep{Kroupa2001,Salpeter1955} $d{\scriptscriptstyle N}/dm=m^{-2.3}$.
The second term is just the MS life time, $t_{{\rm MS}}$, i.e. the
time it takes a MS star to evolve into a WD, $dm/dt=t^{-4/3}$. For
these simplifying assumptions we get 
\begin{equation}
\frac{d{\scriptscriptstyle N}}{dt}\propto t^{2.3/3}t^{-4/3}=t^{-0.56}\approx t^{-3/5}.
\end{equation}

The DTD distribution given here is the predicted DTD of the transients discussed in subsection \ref{sec:WD-MS-collision}. Moreover, 
due to the functional form of the DTD, we expect to find these events not just in star forming spiral galaxies but also in ellipticals.

\section{Discussion and summary}
\label{sec:Discussion-and-summary}


\subsection{Summary}
\label{subsec:Summary}
In this study we described the interactions of triple systems composed
of two equal mass WDs and a low mass MS star with field stars of the
host galaxy. We found that a significant fraction of these systems
become unstable due to interactions with random flyby stars. Those
flybys can excite the outer eccentricity of the triple sufficiently to destabilize
the triple system. The instability manifests itself via multiple
binary-single encounters during which a collision between the MS star
and a WD is plausible. In the case where the triple survives the chaotic
evolution, i.e. there is no collision between any two components, the systems disrupts to 
the final endstate of a compact binary, formed of any two object of the
triple, and an ``escaper'' that is ejected to infinity. The newly formed compact binary can, 
in turn cause a WD-MS collision, or a DWD inspiral
through GW emission, similar to the classic DD scenario that leads
to a Type Ia SN. 

We found that the collision rates are very sensitive to the initial
triple population and its characteristics. For a set of plausible assumptions
we expect a collision rate of $\sim1$ event every $10^{4}{\rm yr}$
in the MW or any similar spiral galaxy, and $\sim1$ event every $5000{\rm yr}$
in an elliptical galaxy with total mass of $10^{11}M_\odot$.

Furthermore, we expect these events to be luminous with $L\approx10^{7}-10^{9}L_{\odot}$ for weeks to years.
We speculate that some sub-luminous SNe or other luminous transients might result. 
Additionally, we calculated the predicted DTD to be $t^{-3/5}$, and note that we expect to
find these events in both star forming galaxies and ellipticals. 
Finally, we estimate that the total rate of these events is of order $1\%$ of the SNIa rate in galaxies.

\section*{Acknowledgments}

E.M. thanks the University of
Maryland CTC prize-fellowship for supporting this
research, and Nathan Leigh for improving this manuscript.

\textbf{Data availability} The data underlying this article will be shared on reasonable request to the corresponding author.

\bibliographystyle{mnras}
\bibliography{TBH.bib}


\end{document}